\documentclass[journal=langd5,manuscript=article]{achemso}

\usepackage[version=3]{mhchem} % Formula subscripts using \ce{}
\usepackage{color}
\author{Mariano Galvagno}
\affiliation{Department of Civil \& Environmental Engineering, Technion, Haifa, Israel}
\author{Guy Z. Ramon}%\email{ramong@technion.ac.il}
\affiliation{Department of Civil \& Environmental Engineering, Technion, Haifa, Israel}
\title[An \textsf{achemso} demo]
  {Hydrodynamic-colloidal interactions of an oil droplet and a membrane surface}
\abbreviations{IR,NMR,UV}

\begin{document}
%%%%%%%%%%%%%%%%%%%%%%%%%%%%%%%%%%%%%%%%%%%%%%%%%%%%%%%%%%%%%%%%%%%%%
%% The manuscript does not need to include \maketitle, which is
%% executed automatically.  The document should begin with an
%% abstract, if appropriate.  If one is given and should not be, the
%% contents will be gobbled.
%%%%%%%%%%%%%%%%%%%%%%%%%%%%%%%%%%%%%%%%%%%%%%%%%%%%%%%%%
%%%%%%%%%%%%
\begin{abstract}
Membranes have been shown to be exceptionally successfully in the challenging separation of stable oil-water emulsions, but suffer from severe fouling that limits their performance. Understanding the mechanisms leading to oil deposition on the membrane surface, as influenced by hydrodynamics and colloidal surface interactions is imperative for informing better engineered membrane surfaces and process conditions. Here, we study the the interactions between an oil droplet and a membrane surface. Hydrodynamics within the water film, confined between the droplet and the membrane, are captured within the framework of the lubrication approximation, coupled with the van der Waals (vdW) and electrostatic interactions through the droplet shape, which is governed by an augmented Young-Laplace equation.  
The model is used to calculate possible equilibrium positions, where the droplet is held at a finite distance from the membrane by a balance of the forces present. 
An equilibrium phase diagram is constructed as a function of  various process parameters, and is shown in terms of the scaled permeation rate through the membrane. The phase diagram identifies the range of conditions leading to deposition, characterized by a `critical' permeation rate, beyond which no equilibrium exists. When equilibrium positions are permitted, we find that these may be classified as  stable/unstable, in the kinetic sense. Further, our results demonstrate the link between the deformation of the droplet and the stability of equilibria. An upward deflection of the droplet surface, owing to a dominant, long-range repulsion, has a stabilizing effect as it maintains the separation between droplet and membrane. Conversely, a downward deflection is de-stabilizing, due to the self-amplifying effect of strongly increasing attractive forces with separation distance - as the surfaces are pulled together due to deformation, the attractive force increases, causing further deformation. This is also manifested by a dependence of the bi-stable region on the deformability of the droplet, which is represented by a capillary number, modified so as to account for the effect of the permeable boundary. As the droplet becomes more easy to deform, the transition from an unconditionally stable region of the phase diagram, to a point beyond which there is no equilibrium (interpreted as deposition) becomes abrupt. These results provide valuable physical insight into the mechanisms that govern oil fouling of membrane surfaces.       
\end{abstract}

\section{Introduction}

The separation of stabilized oil--in--water emulsions poses a difficult technological challenge, often with important environmental implications. 
This is particularly so when treating oily wastewater from various industries, including oil and gas production, prior to discharge so as to minimise pollution and contamination of freshwater sources and the marine environment\cite{Mondal2008,Vengosh2014}. 
Current treatment methods include flotation, coagulation, biological treatment, membrane separation technology, advanced oxidation processes and combined technologies\cite{Ahmadunetal09-jhm,Shafferetal13-envscitech}.  
In particular, membranes have been successful in effectively separating stable emulsions of oil droplets ($<20\mu$m in diameter), difficult to achieve by other techniques \cite{Shafferetal13-envscitech,Tanudjaja2019}. While exceptionally successful at performing the actual separation, membranes suffer from severe fouling due to oil deposition during operation, which results in loss of productivity and requires extensive back--washing and cleaning that can considerably increase costs. 

Fouling is a long-standing issue in membrane separation, particularly when colloidal material is involved. 
In many cases, understanding the characteristics of the specific process, namely the separated mixture, the membrane used and the operating conditions -- particularly the permeation flux through the membrane - can be used to identify a `critical flux', below which fouling is minimized \cite{Bacchin2006}. 
The main idea behind this concept is that the primary cause of colloidal deposition is the permeation through the membrane, so that if some repulsive forces are present, choosing the right permeation rate can reduce deposition significantly. 
Furthermore, in certain cases  deposition has been shown to be reversible -- a particle seemingly deposited at the membrane surface is released upon shutting off of the permeation \cite{Wang2005}. 
Understanding the influence of hydrodynamic force due to permeation, and how it balances against surface interactions (such as electrostatic repulsion) between colloidal particles and membranes will allow for better design of membrane materials and process conditions; this is particularly so for emulsions, where micron-scale droplets are involved. 
While there has been much work devoted to modifying the membrane surface, imparting anti-fouling properties\cite{Zhuetal14-NPGasiamaterials}, there is still insufficient mechanistic understanding of oil droplet deposition, and how this is affected by droplet deformation. 

Recent experimental work has begun to provide insight on droplet behavior at the membrane surface, using microscopic observation\cite{tummonsetal16-jmemsci,fuxetal17-envscitech,Tummons2017,Tanudjaja2018,Tanudjaja2019a}. These have shown various aspects such as droplet accumulation, coalescence and release. 
In particular, it has been shown that there is a link between droplet deformation, as measured using confocal microscopy imaging analysis, and the reversibility of deposition -- droplets that retained a near--spherical shape were easily washed off the membrane upon shutting off of the permeation, while deformed droplets remained attached \cite{fuxetal17-envscitech}. 
%It was shown experimentally that droplets sitting on a membrane deform from an almost perfectly sphere to nearly a hemisphere by increasing the permeate flux and that after subsequently decrease the permeate flux . 

The hydrodynamic interaction between a rigid sphere and a permeable wall has been Theoretically studied quite extensively (the interested reader may find many of these studies summarized in ref \cite{ramonetal13-pof}). 
In particular, the increased viscous drag induced by the proximity to a permeable boundary has been studied in the context of the low permeabilities and colloidal particle sizes representative of commercial membrane separations\cite{Ramonetal12-jmemsci}, and also considered the effect of shape and of the possible existence of equilibrium positions at a finite distance from the membrane surface \cite{ramonetal13-pof}. While providing important insight, however, these studies all consider rigid, non--deformable particles. 
%and hydrodynamic and electrostatic interactions for elastic spheres and a permeable membrane\cite{ramon-notes}. 

Herein, we study the case of a single droplet in equilibrium, at close proximity to a filtration membrane, through which the surrounding fluid flows. 
Specifically, a mathematical model is derived, capturing the interplay between droplet deformation and the resultant forces acting on the droplet due to hydrodynamic and colloidal surface interactions - incorporated via a disjoining pressure.
%We perform numerical simulations varying characteristic process parameters to analyse steady states prescribing a zero net force acting on the droplet to investigate the influence of the particle shape, distance to the membrane, membrane permeability and characteristic parameters of the Derjaguin pressure and EDL contributions on the repulsion / attraction towards the surface. 
% varying the main process parameters. % we derive a set of equations for the gap and pressure in between the deformable axisymmetric droplet and the permeable rigid membrane.
The model is then used to identify the existence of equilibrium positions of the droplet at a finite distance from the membrane surface, the stability of of equilibria and dependence on droplet shape and the various parameters involved. 
%\review{investigate whether there is reversible deposition onto the membrane. We finalise by summarising the main results and future work.}

%%%%%%%%%%%%%%%%%%%%%%%%%%%%%%%%%%%%%%%%%%%%%%%%%%%%%%%%%%%%%%%%%%%%%%%%%%%%%%%%%%%%%%%%%%%%%%%%

\section{Problem formulation}
\subsection{Geometry and long--wave approximation}
\begin{figure}[h!]
\center% \centerline{\includegraphics{Fig1crop}}
\includegraphics[width=0.55\textwidth]{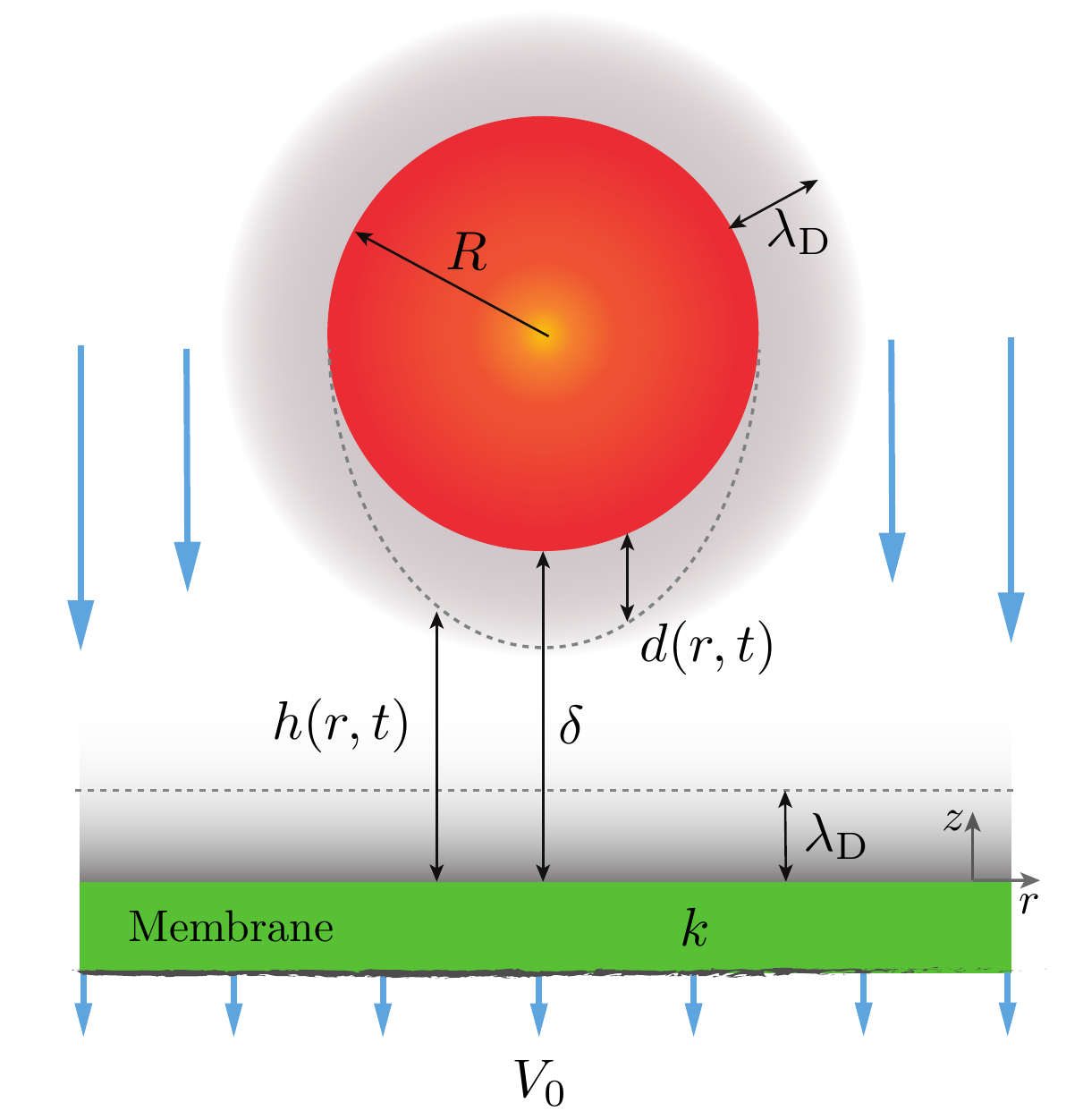}
\caption[Sketch]{Schematic illustration of a liquid droplet with radius $R$, immersed in a liquid close to a membrane with permeability per unit thickness $k$. %$\mathrm{\hat{k}}$.
$h(r,t)\simeq \delta + r^2/2R+d$ is the thickness of the layer confined between the droplet and the membrane, in which the droplet deformation is $d(r,t)$ and $\delta$ is the distance of closest approach between an undeformed droplet and the membrane (note that the deformation is shown to be negative in the sketch, but can be positive as well). The permeation velocity through the membrane is $V_0$ and $\lambda_\mathrm{D}$ is the Debye length.} %\mariano{nicer sketch}}
\label{fig:1}
\end{figure}

We consider an initially spherical oil droplet, with radius $R$, immersed in an incompressible Newtonian fluid, at close proximity to a permeable surface through which a flow is driven (see Fig.~\ref{fig:1} for a schematic illustration of the system). The permeable surface (a separation membrane) is assumed to have a uniform permeance (permeability per unit thickness) $k$, %$\hat{k}$, 
and $V_0$ represents the uniform permeation rate through the membrane, in the absence of the droplet. %and the permeation rate through the membrane is $V_0$. 

%our derivation by following Ramon et al. \cite{ramonetal13-pof, ramon-notes} and Manor et al. \cite{Manoretal08-lang} and references therein.  A sketch of the problem is shown in : 

%%%%%%%%%%%%%%%%%%%%%%%%%%%%%%%%%%%%%%%%%%%%%%%%%%%%%%%%%%%%%%%%%%%%%%%%%%%%%%%%%%%%%%%%%%%%%%%%

%%%%%%%%%%%%%%%%%%%%%%%%%%%%%%%%%%%%%%%%%%%%%%%%%%%%%%%%%%%%%%%%%%%%%%%%%%%%%%%%%%%%%%%%%%%%%%%%

The equations of motion and continuity of the fluid confined within the gap between the approaching droplet and the membrane surface can be significantly simplified by invoking the lubrication approximation, valid when $h<<R$ \cite{oronetal97-rmphys}. Furthermore, we assume that the interface is immobile, corresponding with either a very large viscosity ratio or the presence of sufficient amount of surfactant molecules \cite{Manoretal08-prl}; this results in an imposed no--slip condition and a situation where the flow inside the droplet may be ignored. Under these assumptions, and accounting for the permeation through the boundary, we may write the equation for the pressure within the thin fluid film, separating the droplet and the membrane, as \cite{Ramonetal12-jmemsci,ramonetal13-pof}
\begin{equation}
\begin{aligned}
&\frac{\partial h}{\partial t}=\frac{1}{12 \mu r} \frac{\partial}{\partial r} \left(r h^3\frac{\partial p}{\partial r}\right)-\frac{k}{\mu}p-V_0,
%\mathcal{P}=p-\frac{A_\mathrm{H}}{h^3}+\zeta e^{-h/\lambda_\mathrm{D}}\nonumber,
\label{eq:dim-eq1}
\end{aligned}
\end{equation}
where $\mu$ is the fluid viscosity, and $p$ is the hydrodynamic pressure. This equation describes the deviation of the pressure from the far-field, background pressure away from the drop (see \cite{Ramonetal12-jmemsci,ramonetal13-pof} for further details of this derivation). 
The shape of the droplet near the apex is governed by the linearized, augmented Young-Laplace equation, representing the normal stress balance at the interface \cite{Manoretal08-lang}
\begin{equation}
\begin{aligned}
&\frac{\sigma}{r}\frac{\partial}{\partial r}\left(r\frac{\partial h}{\partial r}\right)=\frac{2\sigma}{R}-\mathcal{P}(r,t).
%\mathcal{P}=p-\frac{A_\mathrm{H}}{h^3}+\zeta e^{-h/\lambda_\mathrm{D}}\nonumber,
\label{eq:dim-eq2}
\end{aligned}
\end{equation}
Here, $\sigma$ is the surface tension coefficient and $\mathcal{P}$ is the generalised stress, defined as
\begin{equation}
\mathcal{P}(r,t)=p+\Pi(h),%\frac{A_\mathrm{H}}{h^3}+\zeta e^{-h/\lambda_\mathrm{D}},
\label{eq:generalised_stress}
\end{equation}
which includes the hydrodynamic pressure $p(r,t)$ and the additional stresses, $\Pi(h)$, resulting from surface interactions (disjoining pressure); here, these are taken as the simple sum of an attractive van--der--Waals stress accounting for the wettability, and a repulsive electrostatic stress \cite{israelachvili2011intermolecular}
%The Derjaguin pressure consists of an attractive colloidal van der Waals potential and a repulsive colloidal electrostatic potential, 
\begin{equation}
\begin{aligned}
\Pi(h)=-\frac{\mathrm{A}_\mathrm{H}}{6\pi h^3}+\zeta e^{-h/\lambda_\mathrm{D}},
\end{aligned}
\end{equation}
%and written more conveniently as
%\begin{equation}
%\begin{aligned}
%\Pi(h)=\zeta\left(-\frac{\mathrm{A}_\mathrm{H}}{6\zeta\pi h^3}+ e^{-h/\lambda_\mathrm{D}}\right),
%\label{eq:disj-pressure-1}
%\end{aligned}
%\end{equation}
in which $\mathrm{\mathrm{A}}_\mathrm{H}$ is the Hamaker constant,  $\lambda_\mathrm{D}$ is the Debye length, representing the characteristic decay length of electrostatic repulsion, and $\zeta$ is a parameter characterising the electrostatic interaction (or electrostatic stress at contact),\cite{israelachvili2011intermolecular} 
\begin{equation}
\zeta=64k_bT c_\infty\tanh\left(\frac{ze\psi_p}{4k_bT}\right)\tanh\left(\frac{ze\psi_m}{4k_bT}\right),
\label{eq:zeta}
\end{equation}
where $k_b$ is the Boltzmann constant, $c_\infty$ denotes the background electrolyte concentration, $z$ the ion valency, $T$ the absolute temperature, $e$ corresponds to the elementary charge and $\psi$ is the electric potential, with subscripts $p$ and $m$ denoting particle and membrane, respectively. 
We note that the choice made here with respect to the colloidal interactions is by no means comprehensive, and mostly serves as an illustrative example of the possible framework offered by the model. For example, more elaborate forms of the electrostatic stress may be used, as well as other forms of the van-der-Waals interaction (e.g., including retardation effects as well as a positive Hamaker constant\cite{israelachvili2011intermolecular,Taboretal11-softmatter}). Certainly, one may prescribe other forms of the disjoining pressure that include structural and solvation interactions, and so forth. 

%It is important to note that the model expressed by equations (1) and (2) assumes a no--slip condition imposed on the surface of the droplet. 
%This represents the physical situation where the presence of a surfactant at the interface immobilises it \cite{Manoretal08-prl}. 
%\review{The choice of a partially wetting disjoining pressure allows for the interplay of }
%
The primary goal of the current study is to examine the stationary droplet, i.e. the case of a droplet at equilibrium. Under such conditions, the net force acting on the droplet must vanish, and is imposed as an integral constraint, 
\begin{equation}
\Sigma\mathrm{F}=\int_0^\infty r\mathcal{P}(r,t)\mathrm{d}r=0.
\label{eq:integral constraint dim}
\end{equation}
Note that the long--wave model formulation focuses on the region of the droplet closest to the membrane, specifically on the gap between the membrane and the droplet -- the lubrication area -- where the hydrodynamic stresses originate. The force calculated from integration of the hydrodynamic stresses does not include the usual `Stokes drag' acting on the entire drop, which has been shown to be much smaller (up to 2-4 orders of magnitude)\cite{Ramonetal12-jmemsci}. We also neglect the effect of deformation on the entire drop, assuming it is confined to a region on the order of ~$\left(Rh\right)^{1/2}$ (for a more detailed view on how stresses change due to whole droplet deformation see ref. \cite{Taboretal11-softmatter, chanetal11-advCollIntSci}).%long-wave approximation and solely focusing on the region around the bottom of the droplet, or more specifically on the gap between the membrane and the droplet ? the lubrication area -, where all the stresses originate. This model does not incorporate deformations of the whole droplet due to the nature of the long-wave approximation area. Future work will extend this model to also take these effects into account.  Deformation over the whole droplet is not n

%\left( p(r,t)-\frac{A_\mathrm{H}}{h(r,t)^3}+\zeta e^{-h(r,t)/\lambda_\mathrm{D}}\right )
\subsection{The scaled, steady-state equation}
To study droplets at equilibrium we solve the steady--state version of Eq.~\eqref{eq:dim-eq1} by setting $\partial h/\partial t =0$. 
Eqs.~\eqref{eq:dim-eq1} and ~\eqref{eq:dim-eq2} 
are non--dimensionalised by scaling the hydrodynamic pressure using a modified viscous stress, $p=(\mu V_0/k)P$, that also incorporates the permeance of the membrane as a length scale. 
Through inspection, balancing the remaining terms in the equations requires the scaling for the radial coordinate $r$ and gap width $h$ to be $r=\eta\left(96kR^3\right)^{1/4}$ and $h=H\left(24kR\right)^{1/2}$, respectively.  %The Disjoining pressure in turn is expressed as $\Pi=\pi_d\tilde{\Pi}$ \cite{ramonetal13-pof}. 
%We consider the droplet to have an unperturbed spherical shape located at a distance $\delta$ from the membrane. 
Using these scaling transformations, we have the steady--state dimensionless equations for the gap width $H(\eta)$ and hydrodynamic pressure $P(\eta)$
\begin{equation}
\begin{aligned}
&\frac{1}{\eta}\frac{\partial}{\partial\eta}\left(\eta H^3\frac{\partial P}{\partial\eta} \right) - P + 1 =0,
%\nonumber
\end{aligned}
\label{eq:thin-film-eq-2a}
\end{equation}
and the scaled Young-Laplace equation
\begin{equation}
\frac{1}{2\eta}\frac{\partial}{\partial\eta}\left(\eta\frac{\partial H}{\partial\eta} \right)-2+\,\widehat{\mathrm{Ca}}\left( P+\frac{1}{\widehat{V}_0}\,\tilde{\Pi}\right)=0.
\end{equation}
Here, as we are also interested in the droplet deformation, we define the gap width as $H=\widehat{\delta}+\eta^2+\widehat{d}$. The deflection $\widehat{d}(\eta)$, and $\widehat{\delta}$, the distance of closest approach to the membrane of an undeformed droplet are scaled against the hydrodynamic decay length $\ell_\mathrm{H}=\left(24kR\right)^{1/2}$. The term $\eta^2$ comes from the parabolic approximation of the unperturbed, spherical droplet shape. Re-casting the steady-state equations in terms of the deflection yields the system
%\begin{tcolorbox}[width=\textwidth,title={{\it System of steady--state scaled evolution equations for $P(\eta)$ and $\widehat{d}(\eta)$ with $\tilde{\Pi}$}}, opacityfill=1]%colback={green},
\begin{equation}
\begin{aligned}
&\frac{1}{\eta}\frac{\partial}{\partial\eta}\left[\eta \left(\widehat{\delta}+\eta^2+\widehat{d}\,\right)^3\frac{\partial P}{\partial\eta} \right] - P + 1 =0,
%&H=\widehat{\delta}+\eta^2+\widehat{d}.
%\nonumber
%\label{eq:thin-film-eq-2}
%\end{aligned}
%\end{equation}
%The total acting force $\widehat{\mathrm{F}}$ should be zero for a droplet in equilibrium, thus the additional scaled integral condition is written as,
%and the additional scaled integral condition 
%\begin{equation}
%\begin{aligned}
%,\mathrm{c}
%\label{eq:integral constraint}
\label{eq:thin-film-eq-2}
\end{aligned}
\end{equation}
\begin{equation}
\frac{1}{\eta}\frac{\partial}{\partial \eta}\left(\eta\frac{\partial\widehat{d}}{\partial\eta}\right)+2\,\widehat{\mathrm{Ca}}\,\left( P+\frac{1}{\widehat{V}_0}\,\tilde{\Pi}\right)=0,
\label{eq:thin-film-eq-3}
\end{equation}
and the scaled equilibrium condition
\begin{equation}
    \widehat{\mathrm{F}}=\int_0^\infty\eta\left(P+\frac{1}{\widehat{V}_0}\tilde{\Pi}\right)\mathrm{d}\eta=0.
\label{eq:thin-film-eq-4}
\end{equation}
%Note that the last term can be either positive or negative, depending on the permeation velocity and droplet velocity. where $H$ is a generic the gap width. 
The scaled equations contain several dimensionless parameters. First, $\widehat{\mathrm{Ca}}=\mu V_0R/\sigma k$ is a modified capillary number, accounting for the ratio of the viscous and the surface tension stresses and differing from the classical capillary number by the factor $R/k$, which comes from the hydrodynamic interaction with the permeable boundary.  
Next, the scaled permeation $\widehat{V}_0=\mu V_0/k\zeta$, represents the ratio of the viscous and repulsive electrostatic stresses at contact. Finally, $\tilde{\Pi}$ is the non--dimensional disjoining pressure 
%%The Derjaguin pressure consists of an attractive colloidal van der Waals potential and a repulsive colloidal electrostatic potential, 
%%\begin{equation}
%%\Pi(h)=-\frac{\mathrm{A}_\mathrm{H}}{6\pi h^3}+\zeta e^{-h/\lambda_\mathrm{D}},
%%\label{eq:disj-pressure}
%%\end{equation}
%%where $\mathrm{\mathrm{A}}_\mathrm{H}$ is the Hamaker constant, $\zeta$ is a parameter characterising the electrostatic interaction (or electrostatic stress at contact)  and $\lambda_\mathrm{D}$ is the Debye repulsion decay length. 
%Eq.~\eqref{eq:disj-pressure-1} is written for simplification as
%\begin{equation}
%\Pi(h)=\pi_d\tilde{\Pi}.
%\label{eq:disj-pressure-simp}
%\end{equation}
%Note that $\tilde{\Pi}$ is non--dimensional and is 
defined as
\begin{equation}
\tilde{\Pi}(H)=-\frac{\widehat{\mathrm{A}}_\mathrm{H}}{H^3}+e^{-H/\widehat{\lambda}_\mathrm{D}},
\label{eq:disj-pressure-simp}
\end{equation}
with $\widehat{\mathrm{A}}_\mathrm{H}=\mathrm{A}_\mathrm{H}/6\pi\zeta\,\ell_\mathrm{H}^{\,3}$ the scaled Hamaker constant, accounting for the ratio of attraction and repulsion stresses and $\widehat{\lambda}_\mathrm{D}=\lambda_\mathrm{D}/\ell_\mathrm{H}$ is the ratio of the electrostatic and the hydrodynamic decay lengths.  %$\delta_0$ is an arbitrary cut-off distance for the vdW interaction. 
%The dimensional constant $\pi_d$ is in turn simply $\pi_d=\zeta$. %, see \cite{ramon-notes} Eq.~(2.13),
%\begin{equation}
%\zeta=64k_bTc_\infty\tanh(\frac{ze\Psi_p}{4k_bT})\tanh(\frac{ze\Psi_w}{4k_bT}).
%\label{eq:zeta}
%\end{equation}
%{\it maybe further considerations regarding colloidal interpretation and citations, relate to membrane etc.}
Typical physical values and ranges of process parameters are shown in Table~\ref{tab:dimesional ranges}, while in 
Table~\ref{tab:non--dim:order:description} we summarise all the non--dimensional parameters and corresponding orders of magnitudes used in the forthcoming analysis.
%\begin{table}[h]
%\begin{center}
%\begin{tabular}{cc|cc|cc}
%\hline 
%&&                                              &&                                 \\
%$\widehat{V}_0=\mu V_0/k\zeta$                                                                                                && $\widehat{\mathrm{Ca}}=\mu V_0R/\sigma k$  && $\widehat{\lambda}_\mathrm{D}=\lambda_{\mathrm{D}}/\ell_{\mathrm{H}}$ \\
%&&                                                                            &&                                                                                                         \\
%$\widehat{\mathrm{A}}_{\mathrm{H}}=\mathrm{A}_{\mathrm{H}}/\zeta\ell_{\mathrm{H}}$        && $\widehat{\delta}=\delta/\ell_{\mathrm{H}}$       &&                  $\ell_{\mathrm{H}}=\left(24kR\right)^{1/2}$                     \\       
%&&&&\\                                                                                               
%%\hline  
%\caption{Non--dimensional parameters of the problem}% (for simplicity units are omitted).}
%\label{tab:non--dimorderofmagnitude}
%\end{tabular}
%\end{center}
%\end{table}

%%%%%%%%%%%%%%%%%%%%%%%%%%%%%%%%%%%%%%%%%%%%%%%%%%%%%%%%%%%%%%%%%%%%%%%%%%%%%%%%%%%%%%%%%%%%%%%%

\begin{table}[]
\footnotesize
\begin{tabular}{l|l|l|l}
% \hline
%\multicolumn{4}{c}{Orders of magnitude for dimensional parameters} \\
%\hline
 \multicolumn{1}{c}{\it{Parameter}} & \multicolumn{1}{c}{\it{Description}} & \multicolumn{1}{c}{\it{Parameter}} & \multicolumn{1}{c}{\it{Description}}\\
%\hline
&&&\\
$\mathrm{A}_{\mathrm{H}}\sim10^{-21}\mathrm{J\,/m\,s^2}$          &  Hamaker constant                            &    $R\,\,\sim10^{-7} - 10^{-5}\mathrm{m}$                                     &  Droplet radius        \\ 
$\zeta\,\,\,\,\,\sim10^{4}\,\mathrm{Pa}$                                    &   Electrostatic stress at contact          &    $\mu\,\,\,\sim10^{-3}\,\mathrm{Pa,s}$                                   &  Viscosity               \\
$\lambda_\mathrm{D}\sim10^{-9} - 10^{-7}\mathrm{m}$                                   &   Debye length                                   &    $V\,\,\sim10^{-5}-10^{-4}\,\mathrm{m/s}$                                               &  Permeation velocity    \\
$\delta\,\,\,\,\,\sim10^{-5}\,\mathrm{m}$                                                &   Distance to the membrane              &    $k\,\,\,\,\sim10^{-13} - 10^{-12}\mathrm{m}$                                &  Membrane permeance \\
$\sigma\,\,\,\,\sim10^{-2}\,\mathrm{N/m}$                                         &   Surface tension                                &    $\ell_{\mathrm{H}}\,\sim10^{-10}-10^{-8}\mathrm{m}$            &   Hydrodynamic decay \\ 
%\hline
\end{tabular}
\caption{Orders of magnitude for dimensional parameters of the problem.}
\label{tab:dimesional ranges}
\end{table} 

%%%%%%%%%%%%%%%%%%%%%%%%%%%%%%%%%%%%%%%%%%%%%%%%%%%%%%%%%%%%%%%%%%%%%%%%%%%%%%%%%%%%%%%%%%%%%%%%

Finally, we specify the boundary conditions imposed on the system of equations.  At the origin, $\eta=0$,  we have symmetry considerations, i.e. 
\begin{equation}
\begin{aligned}
\frac{\partial\widehat{d}}{\partial\eta}&=0,\\%\text{\,\,\,\,\,\,\,\,\,at\,\,\,$\eta=0$}\\
\frac{\partial P}{\partial\eta}&=0.%\text{\,\,\,\,\,\,\,\,\,at\,\,\,$\eta=0$},
\end{aligned}
\label{eq:BC1a}
\end{equation}
Far from the apex, we expect the pressure to decay back to the background value, and the deflection to likewise vanish\cite{YiantsiosDavis90-jfm, Manoretal08-lang, taboretal12-jcolIntSci} as $\eta\rightarrow\infty$ which, for the numerical scheme corresponds to the simulation domain $\eta=L$, so we impose
\begin{equation}
\begin{aligned}
\widehat{d}&=0\\%(\eta\rightarrow\infty)=0\\
%\partial_\eta\widehat{d}(\eta\rightarrow\infty)&=0\\
%&P(\eta\rightarrow\infty)=0.
\frac{\partial P}{\partial\eta}+4\frac{P}{\eta} &=0.
\end{aligned}
\label{eq:BC1b}
\end{equation}
%%%%%%%%%%%%%%%%%%%%%%%%%%%%%%%%%%%%%%%%%%%%%%%%%%%%%%%%%%%%%%%%%%%%%%%%%%%%%%%%%%%%%%%%%%%%%%%%

%The condition of vanishing pressure at $\infty$  can also be expressed as
%$\partial_\eta P+4/\eta P=0$ \cite{YiantsiosDavis90-jfm}. 

%\begin{table}[]
%\begin{tabular}{l|l|l|l}
%\hline
%\multicolumn{4}{c}{Orders of magnitude for dimensional parameters} \\
%\hline
% &&\\
% $\mathrm{A}_{\mathrm{H}}\sim10^{-21}\mathrm{kg\,/m\,s^2}$  &    $\zeta\sim10^{4}\,\mathrm{kg\,/m\,s^2}$  & $\lambda_\mathrm{D}\sim10^{-9} - 10^{-7}\mathrm{m}$  & $\mu\sim10^{-3}\,\mathrm{kg/m\,s}$       \\ 
% $R\sim10^{-7} - 10^{-5}\mathrm{m}$                                          &    $k\sim10^{-13} - 10^{-12}\mathrm{m}$  &  $V\sim10^{-4}\,\mathrm{m/s}$                            &  $\sigma\sim10^{-2}\,\mathrm{kg/s^2}$    \\
% $\delta\sim10^{-5}\,\mathrm{m}$                                                &     $\ell_{\mathrm{H}}\sim10^{-10}-10^{-8}\mathrm{m}$                                                            &                                                                   \\
% &&
%\caption{orders of magnitude for dimensional parameters of the problem.}
%\label{tab:dimesional ranges}
%\end{tabular}
%\end{table} 
%%%%%%%%%%%%%%%%%%%%%%%%%%%%%%%%%%%%%%%%%%%%%%%%%%%%%%%%%%%%%%%%%%%%%%%%%%%%%%%%%%%%%%%%%%%%%%%%

\begin{table}
    \footnotesize
    \begin{tabular}{l|r|l}
     %   \hline
       % \multicolumn{1}{c|}{}                                                                                                   &  \multicolumn{1}{c|}{}          &  \multicolumn{1}{c}{}\\
        \multicolumn{1}{c}{\it{Non--dimensional parameters}}                                                                           &  \multicolumn{1}{c}{\it{Characteristic}}                                                          &  \multicolumn{1}{c}{\it{Description}}\\
        \multicolumn{1}{c}{}                                                                                                                                &  \multicolumn{1}{c}{\it{ ranges}}                                                                    &  \multicolumn{1}{c}{}\\
       %\hline
        && \\
        $\widehat{V}_0=\mu V_0/k\zeta$                                                                                                            & $10^{-4}-10^4$                                                                                            & Ratio of viscous--repulsive stresses                \\ 
        &&\\
        $\widehat{\mathrm{Ca}}=\mu V_0R/\sigma k$                                                                                       & $0-100$                                                                                                        & Ratio of viscous--surface tension stresses      \\ 
        &&\\
        $\widehat{\lambda}_\mathrm{D}=\lambda_{\mathrm{D}}/\ell_{\mathrm{H}}$                                                         & $1-100$                                                                                                        & Ratio of electrostatic--hydrodynamic               \\ 
        && decay length scales   \\
        $\widehat{\mathrm{A}}_{\mathrm{H}}=\mathrm{A}_{\mathrm{H}}/6\pi\zeta\ell^3_{\mathrm{H}}$            & $10^{-3}-1$                                                                                                   & Scaled colloidal stress                                      \\ 
        &&\\
        $\,\,\,\,\,\widehat{\delta}=\delta/\ell_{\mathrm{H}}$                                                                                   & $10^{-5}-10^3$                                                                                              & Scaled distance of closest approach               \\
            &&\\
        $\ell_\mathrm{H}\,\,=\left(24kR\right)^{1/2}$                                                                                            &     $10^{-10}\mathrm{m}-10^{-8}\mathrm{m}$                                                & Hydrodynamic decay length                                      \\
      %&&\\
%\hline
     \end{tabular} %\hline
        % &&\\
%        $\ell_\mathrm{H}=\left(24kR\right)^{1/2}$                                                                                       &     $10^{-10}\mathrm{m}-10^{-8}\mathrm{m}$                                                & Hydrodynamic length                                      \\
%       &&% \hline
       \caption{Definition of non--dimensional parameters, characteristic ranges of orders of magnitude and description. The hydrodynamic decay length is defined as $\ell_\mathrm{H}=\left(24kR\right)^{1/2}$. }
     \label{tab:non--dim:order:description}
\end{table}

%where $\mathrm{c}$ is an arbitrary constant.
%(or for $\eta\gg\eta_{max})$ 

%%%%%%%%%%%%%%%%%%%%%%%%%%%%%%%%%%%%%%%%%%%%%%%%%%%%%%%%%%%%%%%%%%%%%%%%%%%%%%%%%%%%%%%%%%%%%%%%

\section{Results and discussion}
In order to obtain the droplet shape and pressure profiles at equilibrium, where the droplet is stationary and under a zero net force, we solve the second--order problem given by Eqs.~\eqref{eq:thin-film-eq-2}-\eqref{eq:thin-film-eq-4} %and %\eqref{eq:integral constraint} ,
along with the boundary conditions presented in Eqs.~\eqref{eq:BC1a} and \eqref{eq:BC1b}. 
The system is solved numerically using the auto07p continuation package\cite{Doedeletal91-intjbifchaos, Dijkstraetal14-ccp}, for parameter ranges described in Table~\ref{tab:non--dim:order:description}. 
For all numerical  calculations, domain size is set to $L=10$, which was found adequate in assuring that the pressure and deformation decay to zero in the far--field, independent of the choice of domain size. 

\subsection{The equilibrium phase diagram}
The main outputs of these calculations are the distributions of the various stress components, in particular the hydrodynamic pressure, as well as the shape of the droplet. 
However, an even more interesting outcome is the very existence of a solution for which an equilibrium exists and $H>0$; beyond a particular region of parameter space, no such equilibrium exists. 
We further find that, for a certain range of parameters, two solutions exist. 
This behaviour was previously described by Ramon et al. \cite{ramonetal13-pof} for rigid spherical particles, but is here modified by the deformation of the droplet shape and the inclusion of the van--der--Waals force.  

\begin{figure}[h!]
\includegraphics[width=0.75\textwidth]{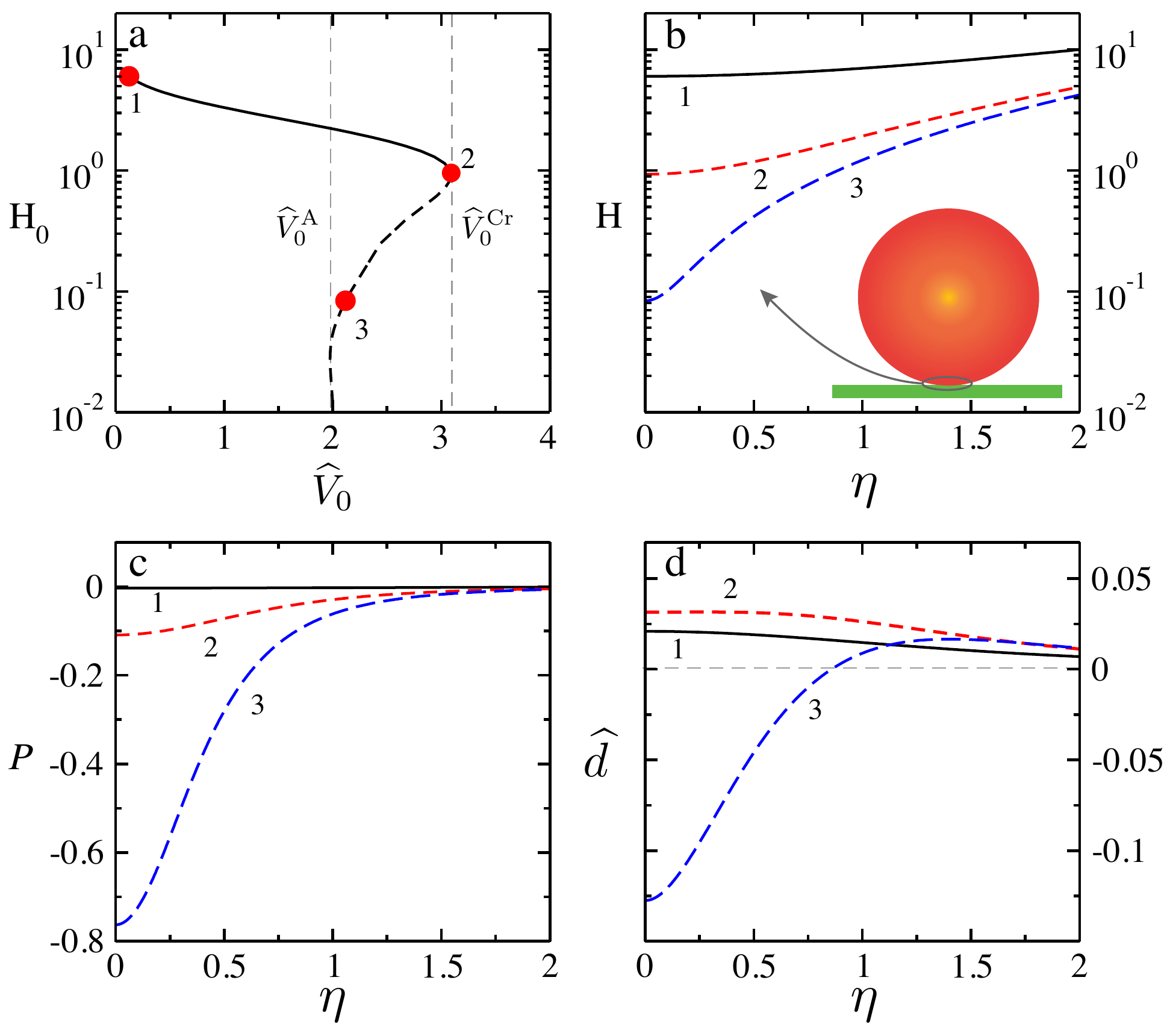}%{drop_sketch}
\caption{Pressure and droplet profiles for different values of modified permeation $\widehat{V}_0$. Panel (a) shows $\mathrm{H_0}$ as a function of the modified permeation $\widehat{V}_0$ for $\widehat{\mathrm{A}}_\mathrm{H}=0.001$, $\widehat{\lambda}_\mathrm{D}=1$ and $\widehat{\mathrm{Ca}}=1$. Labels correspond to droplet profiles, pressure distribution and deflection profiles shown in subsequent panels. Dashed black line corresponds to unstable solutions branch. Panel (b) shows droplet profiles for different values of $\widehat{V}_0$ as indicated. The inset indicates the region of interest. Panels (c) and (d) depict pressure distribution and deflection profiles respectively.}
\label{fig:2}
\end{figure}

The measure used to construct the phase diagrams is the distance between the droplet and the membrane at the origin, $\mathrm{H}_0\equiv H(0)=\widehat{\delta}+\widehat{d}(0)$, plotted against the scaled permeation, $\widehat{V}_0\equiv \mu V_0 R/\sigma k$, which represents a main feature of the current problem -- the permeable boundary, a defining characteristic of the separation membrane (see Fig.~\ref{fig:2}a for an example of the phase diagram and its general features). 
When a finite distance separates the droplet from the membrane under equilibrium, it means that adhesion may be prevented by repulsive forces. 
This distance would be smaller or larger than that obtained for a rigid particle, dependent on whether there is a downward or upward deflection of the droplet surface, respectively.
When no equilibrium solution exists we interpret this as deposition -- the droplet makes contact with the surface. 
Finally, when two solutions exist, one solution is understood to be stable, at least in the kinetic sense, while the other is unstable. Kinetic stability refers, here as in the classical sense, to the existence of an energy barrier in the presence of Brownian motion; even if the force balance predicts an equilibrium position, there may still be a thermal `kick' large enough to overcome the energy barrier and cause the surfaces to make contact. 
We note that the calculation of the energy barrier and, hence, a measure of the actual kinetic stability and its characteristic time--scale, requires the solution of the full transient problem and is beyond the scope of the present study. 
The point of vanishing stable solutions is also where the unstable branch emerges. 
On a plot of $\mathrm{H}_0$ vs. $\hat{V}_0$, this point (marked as point 2 on Fig.~\ref{fig:2}a) embodies the existence of the `critical flux', $\widehat{V}_0^{Cr}$, for a given membrane--emulsion system, as beyond this point deposition will always occur. 
Since $\widehat{V}_0$ represents the operating permeation rate and properties of the emulsion, it allows a choice of operating conditions to shift the system from regions of rapid deposition to regions of delayed deposition. 
%%%%%%%%%%%%%%%%%%%%%%%%%%%%%%%%%%%%%%%%%%%%%%%%%%%%%%%%%%%%%%%%%%%%%%%%%%%%%%%%%%%%%%%%%%%%%%%%

\subsubsection{Droplet profiles at equi--valued scaled permeation}

\begin{figure}[h!]
\includegraphics[width=0.75\textwidth]{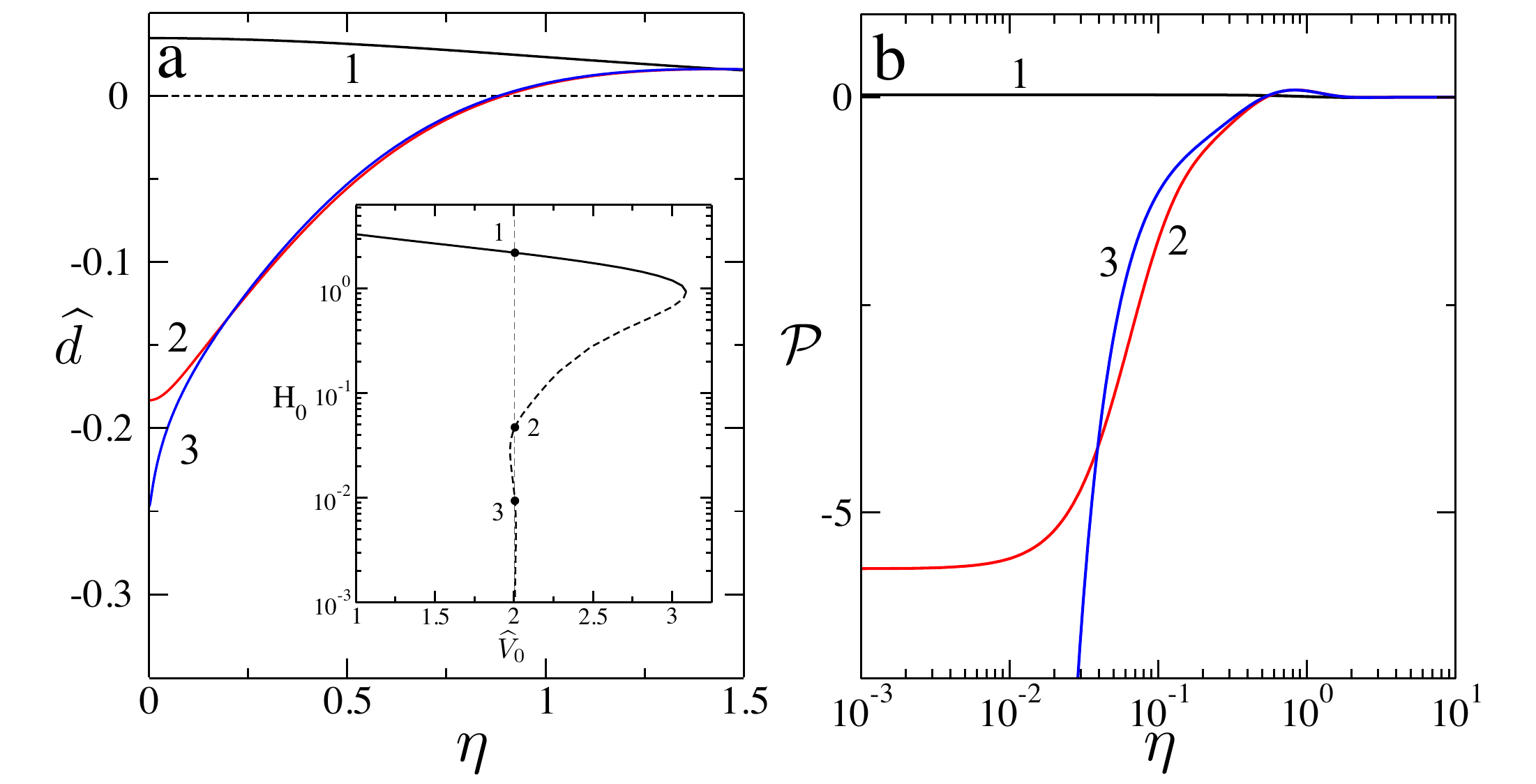}

\includegraphics[width=0.75\textwidth]{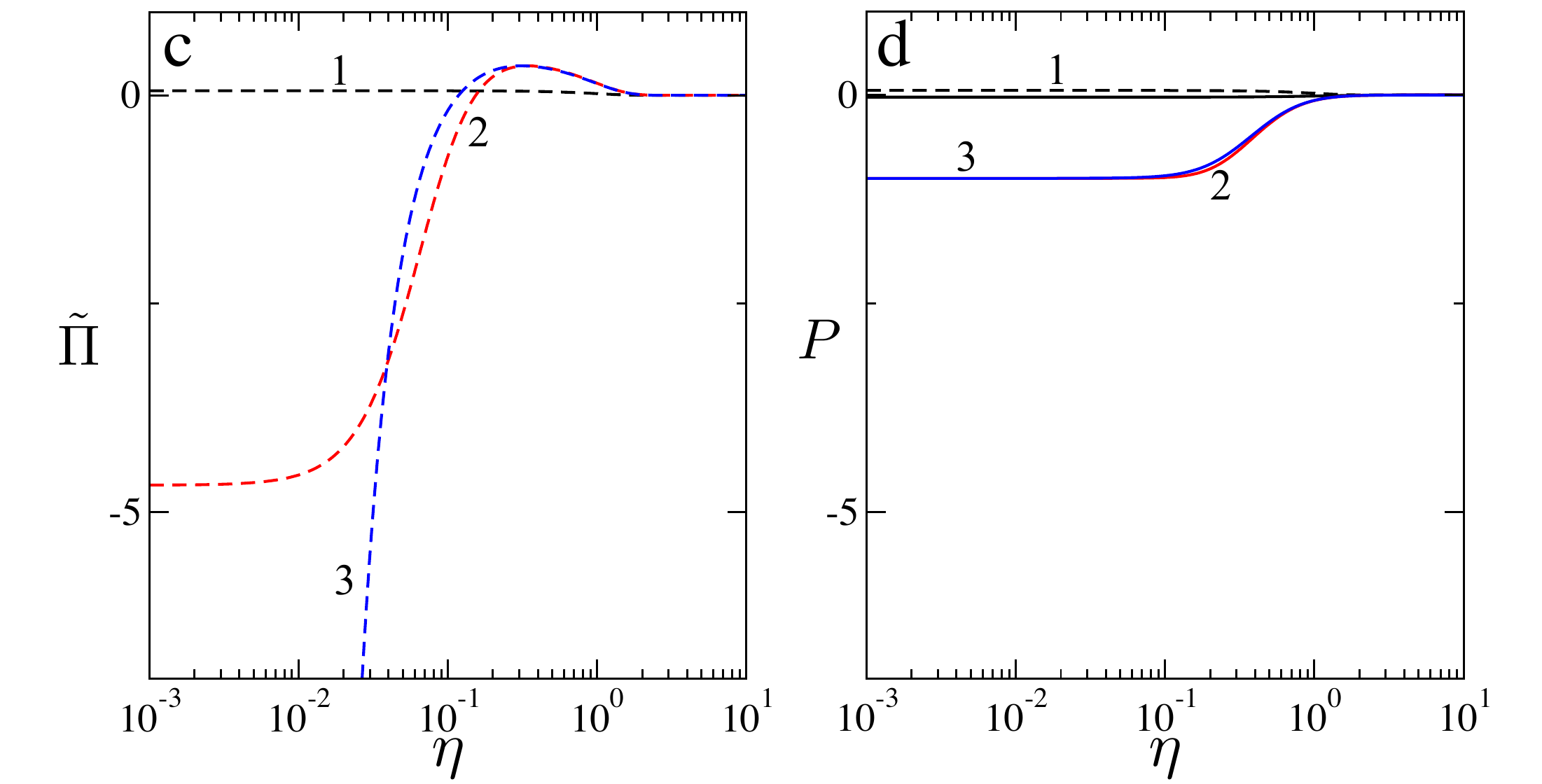}
\caption{Deflection and stress profiles for solutions to equal--values scaled permeation $\widehat{V}_0\sim2$, for $\widehat{\mathrm{A}}_\mathrm{H}=0.001$, $\widehat{\lambda}_{D}=1$, $\widehat{\mathrm{Ca}}=1$, corresponding with the phase diagram shown in the inset of panel (a), in which labels 1--3 correspond to profiles shown in subsequent panels. (a) the deflection profiles $\widehat{d}$, note the steepness of the deflection in profile (3). (b) the generalised stress $\mathcal{P}$ (c) the colloidal stress $\tilde{\Pi}$ (d) the hydrodynamic stress $P$.} %
\label{fig:3}
\end{figure}

Interesting features that accompany the equilibrium solution are the trends in the distributions of the pressure and deflection, as well as the overall droplet shape near the origin. 
In order to further understand this behavior, we examine the case of solutions found for an equal value of the scaled permeation rate, $\hat{V}_0\sim2$, and their differences. 
%Following Fig.~\ref{fig:2}a, points 1--3 marked on the phase diagram signify, on each of the subsequent plots, stable vs. unstable solutions and their respective pressure, deflection and shape profiles. 
Following the inset of Fig.~\ref{fig:3}a, points 1--3 marked on the phase diagram signify, on each of the subsequent plots, stable vs. unstable deflections (a) and their corresponding generalised stress (b), colloidal stress (c) and hydrodynamic stress (d) profiles.
Stable solutions are seen to be upward-deflecting, meaning that repulsion is significant enough to push the droplet surface away from the membrane surface, resulting in a stable solution -- no adhesion. 
Conversely, unstable solutions are seen to be downward--deflected, which reduces the gap between the droplet and membrane surfaces compared with the equivalent, rigid case. 
The reason behind the unstable nature of this solution lies in the physics of the hydrodynamic interaction, which is the main attractive force acting on the droplet at longer ranges. 
This interaction increases as the separation distance decreases, so a downward--deflection is a self--amplifying mechanism -- the permeation decreases the pressure in the confined gap between the two surfaces, which causes the downward deflection, which further decreases the pressure and so on. 
The scales for both the hydrodynamic stress and the scaled colloidal stress show that the attractive colloidal stress component becomes stronger than the electrostatic repulsion as the droplet apex gets closer to the membrane, and thus increases the negative deflection. 
This presumably promotes the irreversible deposition of the droplet on the membrane. 
The case examined shows the existence of a stable profile (1), and two unstable profiles, (2) and (3) for the same modified permeation $\widehat{V}_0$. So we find that, compared with the behavior of a rigid particle, deformability can have a stabilizing effect, but then also exhibits a more abrupt transition. 
The cusping is due to van--der--Waals attraction, that become dominant at close proximity and eventually induces a profile reminiscent of `pinch--off' at the droplet leading edge.  

%%%%%%%%%%%%%%%%%%%%%%%%%%%%%%%%%%%%%%%%%%%%%%%%%%%%%%%%%%%%%%%%%%%%%%%%%%%%%%%%%%%%%%%%%%%%%%%

%%%%%%%%%%%%%%%%%%%%%%%%%%%%%%%%%%%%%%%%%%%%%%%%%%%%%%%%%%%%%%%%%%%%%%%%%%%%%%%%%%%%%%%%%%%%%%%%

%\clearpage
\subsection{Influence of process parameters on equilibria and the `critical flux'}

%%%%%%%%%%%%%%%%%%%%%%%%%%%%%%%%%%%%%%%%%%%%%%%%%%%%%%%%%%%%%%%%%%%%%%%%%%%%%%%%%%%%%%%%%%%%%%%%

\begin{figure}[ht!]
\includegraphics[width=0.49\textwidth]{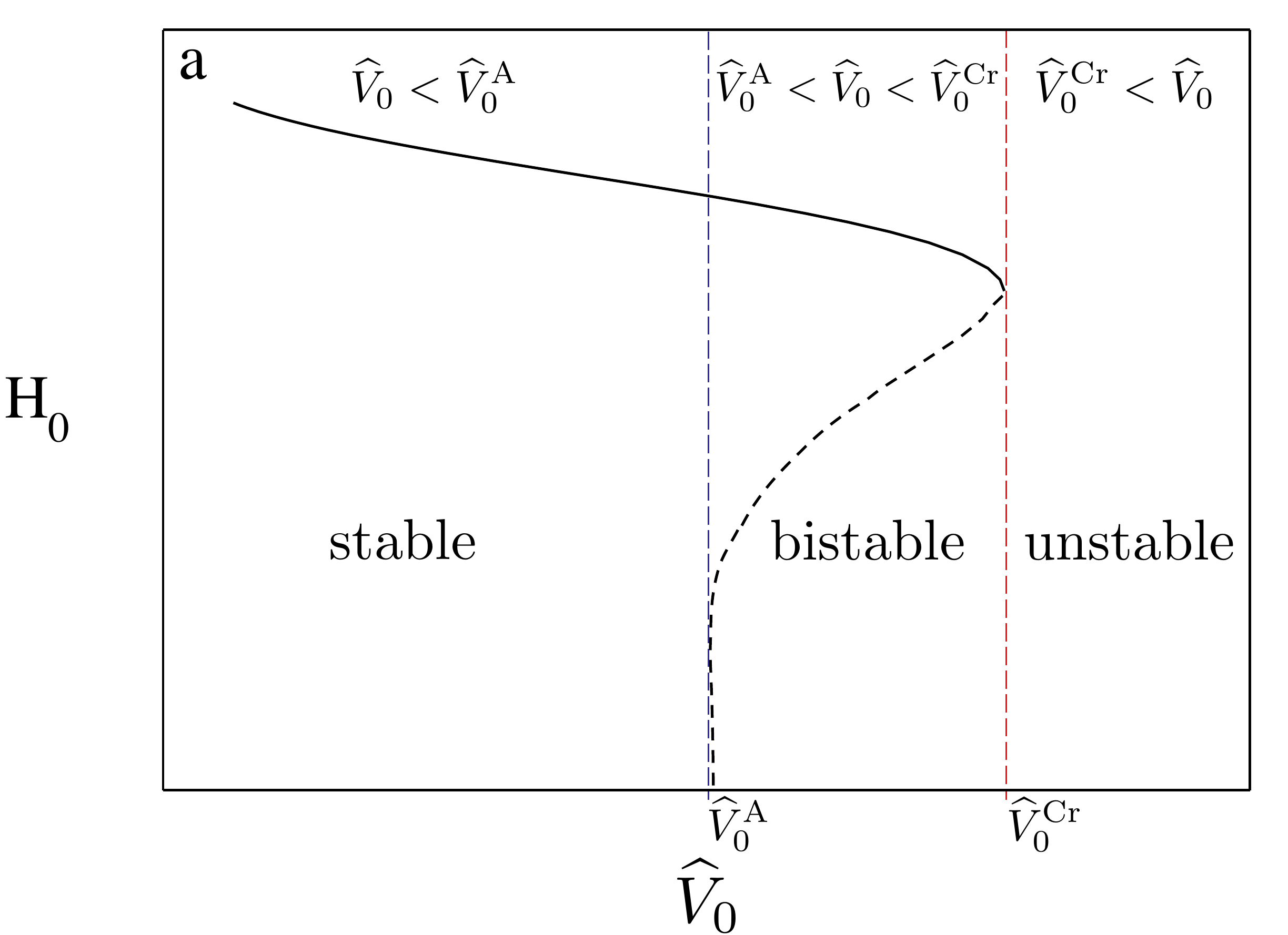}%{drop_sketch}
\includegraphics[width=0.49\textwidth]{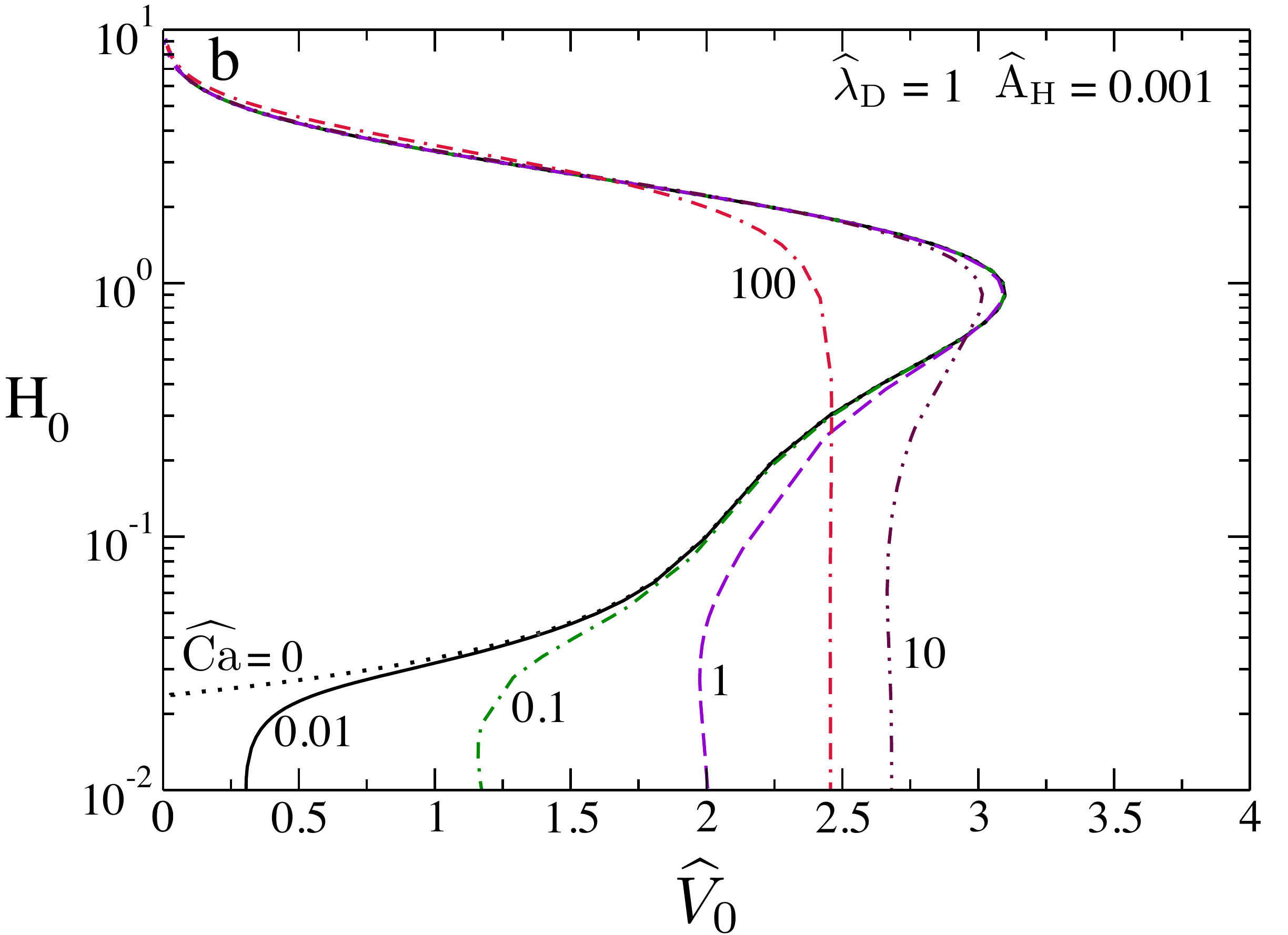}%{drop_sketch}

\includegraphics[width=0.49\textwidth]{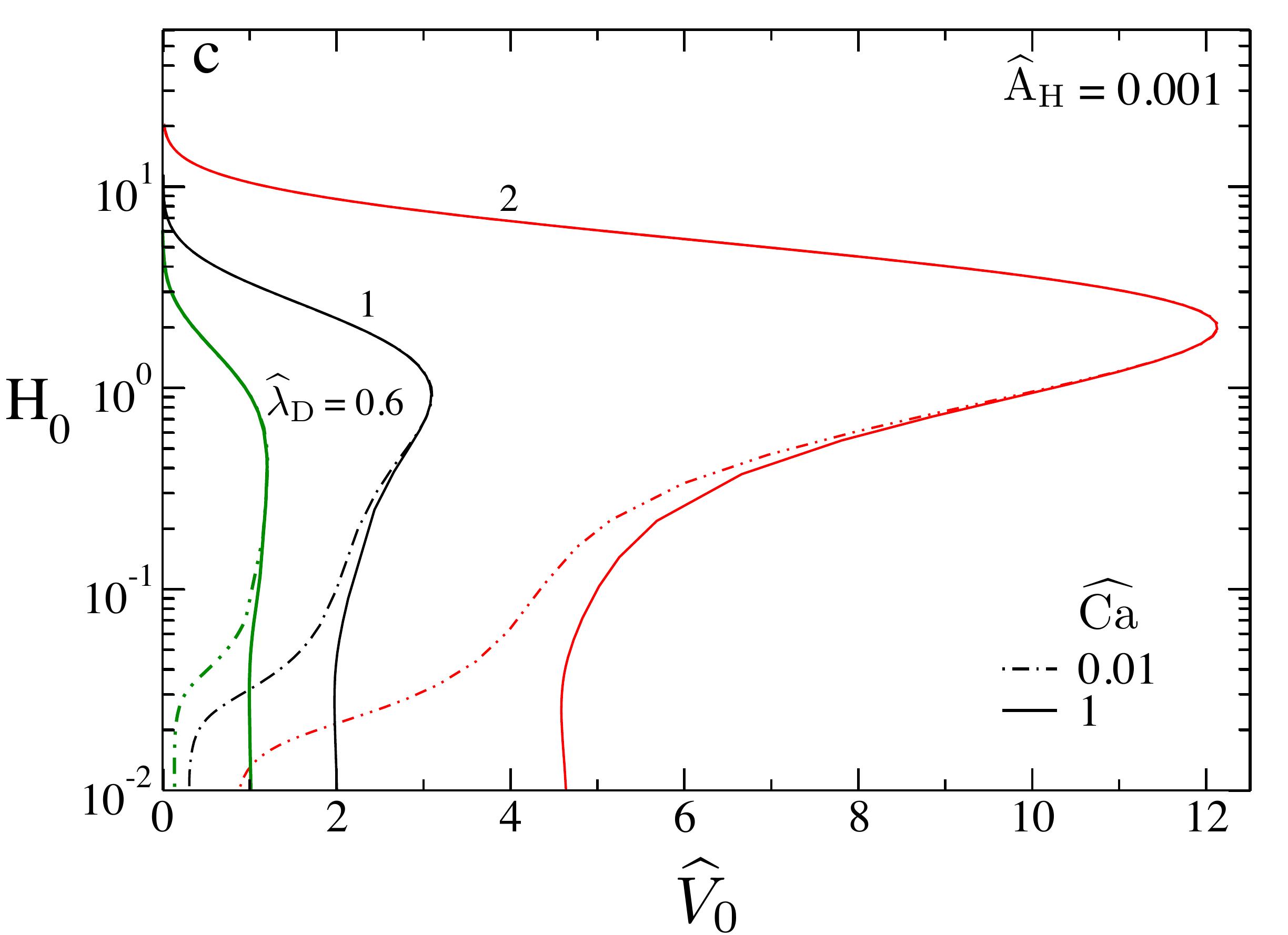}%f}%{drop_sketch}
\includegraphics[width=0.49\textwidth]{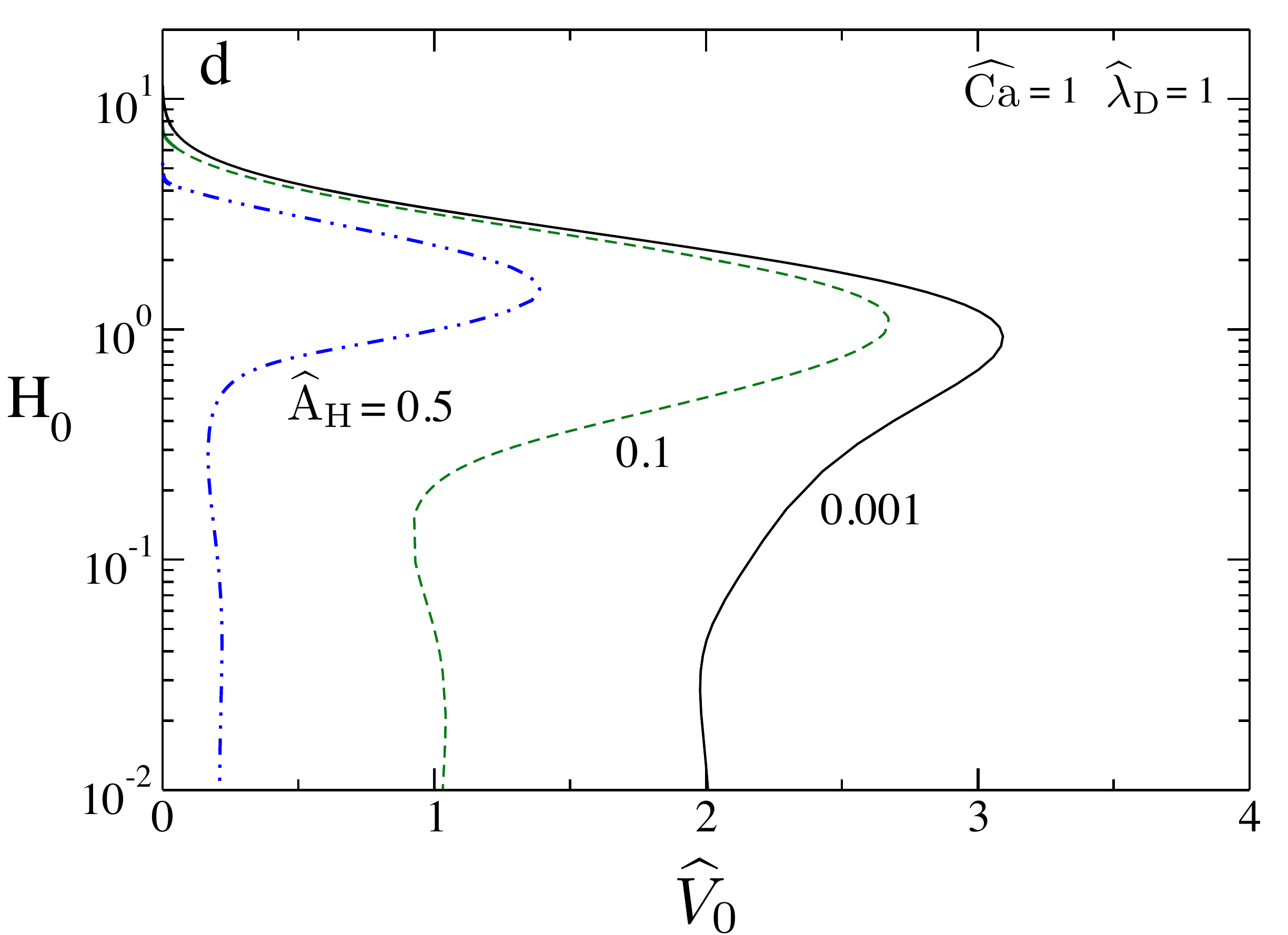}%{drop_sketch}
\caption[Sketch]{(a) A sketch of the equilibrium phase diagram, identifying the critical scaled permeation $\widehat{V}_0^{\mathrm{Cr}}$ and the detachment permeation $\widehat{V}_0^{\mathrm{A}}$, as well as different stability regions. Upper branch correspond to stable solutions (solid black line) and lower branch to unstable solutions (dashed line). (b)  The gap width at the origin $\mathrm{H_0}$ as a function of the scaled permeation $\widehat{V}_0$ for different values of the modified capillary number $\widehat{\mathrm{Ca}}\equiv \mu V R/\sigma k$. The scaled Debye length $\widehat{\lambda}_\mathrm{D}=1$ and Hamaker constant $\widehat{\mathrm{A}}_\mathrm{H}=0.001$. %As the modified capillary number $\widehat{\mathrm{Ca}}$ increases the detachement permeation $\widehat{V}_0^{\mathrm{A}}$ increases. Note that the critical value of $\widehat{V}_0^{\mathrm{Cr}}$ remains the same until the hydrodynamic stresses become stronger and the abrupt detachment  permeation $\widehat{V}_0^{\mathrm{A}}$ starts to decrease. 
(c) $\mathrm{H_0}$ vs. $\widehat{V}_0$ for different values of the modified capillary number $\widehat{\mathrm{Ca}}$ and scaled Debye length $\widehat{\lambda}_\mathrm{D}$. 
%As $\widehat{\lambda}_\mathrm{D}$ increases, both the critical modified permeation  $\widehat{V}_0^{\mathrm{Cr}}$ and  detachment permeation $\widehat{V}_0^{\mathrm{A}}$ increase. Panel 
(d) $\mathrm{H_0}$ vs. $\widehat{V}_0$ for different values of the scaled Hamaker constant $\widehat{\mathrm{A}}_\mathrm{H}=0.001,0.1$ and $0.5$.  
%As $\widehat{\mathrm{A}}_\mathrm{H}$ increases, both the critical modified permeation  $\widehat{V}_0^{\mathrm{Cr}}$ and  detachment permeation $\widehat{V}_0^{\mathrm{A}}$ decrease.
} 
\label{fig:4}
\end{figure}
%%%%%%%%%%%%%%%%%%%%%%%%%%%%%%%%%%%%%%%%%%%%%%%%%%%%%%%%%%%%%%%%%%%%%%%%%%%%%%%%%%%%%%%%%%%%%%%%

The influence of the various parameters characterizing the process is illustrated in Fig.~\ref{fig:4}. 
specifically, the deformability of the droplet is governed by the modified capillary number $\widehat{\mathrm{Ca}}$, representing the ratio of viscous forces tending to deform the droplet and surface tension that tends to retain the spherical shape (see for example   ref.\cite{taboretal12-jcolIntSci}); the scaled Debye length $\widehat{\lambda}_{\mathrm{D}}$ represents the ratio of electrostatic--hydrodynamic decay lengths and hence their relative dominance at long--range; finally, the scaled Hamaker constant  $\widehat{\mathrm{A}}_\mathrm{H}$ indicates the ratio of attractive--repulsive colloidal stresses considered in the current problem. 
A sketch of the equilibrium phase diagram and corresponding equilibria regions is presented in  Fig.~\ref{fig:4}a, separating regions of stable and unstable parameter space. 
As already mentioned earlier, for a given set of parameters there is a value of $\widehat{V}_0$ above which no equilibrium exists and this is interpreted as deposition of the droplet onto the membrane, occuring beyond a `critical' permeation ($\widehat{V}_0^{\mathrm{Cr}}$). 
However, we also distinguish between two regions that do permit equilibria -- one region in which both a stable and an unstable solution exist (for $\widehat{V}_0^{\mathrm{A}}<\widehat{V}_0<\widehat{V}_0^{\mathrm{Cr}}$, where we define $\widehat{V}_0^{\mathrm{A}}$ as the point where the unstable branch corresponds with 'pinch-off' of the droplet leading edge), and another which is unconditionally stable. 
The latter appears to be the consequence of the droplet deformability, as shown in Fig.~\ref{fig:4}b. For a rigid particle ($\widehat{\mathrm{Ca}}=0$), in the presence of vdW attraction, such an unconditionally stable region does not exist. 
However, we see that as $\widehat{\mathrm{Ca}}$ increases, indicating a stronger tendency of the droplet to deform, two things occur; first, the unconditionally stable region is pushed to higher permeation rates. 
This is presumably the consequence of strongly repulsive conditions, under which the the droplet experiences an upward deflection and does not make contact with the membrane, and this tendency increases as it becomes easier to deform the droplet. 
The second noticeable effect is that the bi--stable region becomes smaller, not only because of the stabilizing effect of deformation, but also, at large enough $\widehat{\mathrm{Ca}}$, since the `critical flux' is decreased -- and so deformation becomes de--stabilizing. 

A reduction of $\widehat{V}_0^{\mathrm{Cr}}$ also occurs when the scaled Debye length is decreased (see panel Fig.~\ref{fig:4}c), which results in a shorter-ranged electrostatic repulsion, compared with the attractive force resulting from the hydrodynamic interaction. Similarly, and as can be expected, a larger scaled Hamaker constant likewise decreases the critical scaled flux (see Panel Fig.~\ref{fig:4}d).   

\begin{figure}[ht!]
\includegraphics[width=1\textwidth]{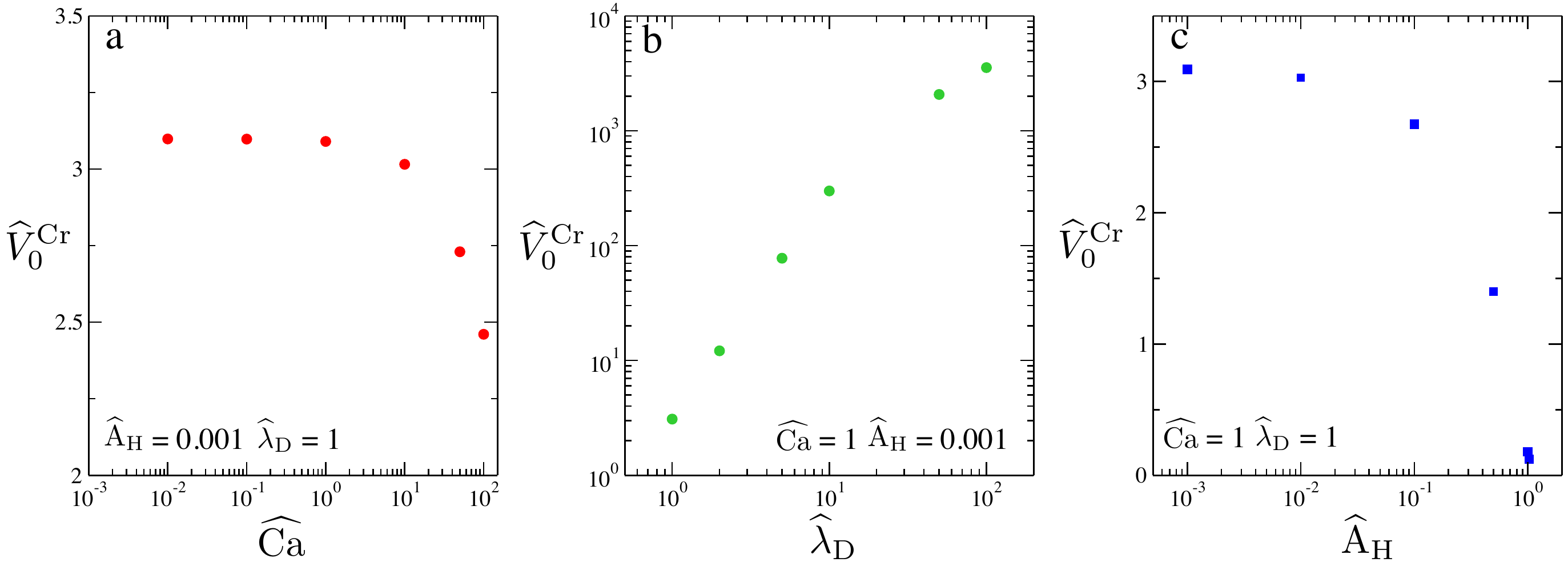}%{drop_sketch}
\caption{Critical permeation flux for different process parameters: The panels depict how the critical permeation flux $\widehat{V}_0^{\mathrm{Cr}}$ change for different values of process parameters:  $\widehat{\mathrm{Ca}}$ (panel (a)), $\widehat{\lambda}_\mathrm{D}$ (panel (b)) and  $\widehat{\mathrm{A}}_\mathrm{H}$ (panel (c)) as shown. }
\label{fig:5}
\end{figure}
%\\
The overall trend observed for the critical scaled flux, $\widehat{V}_0^{\mathrm{Cr}}$, is shown as a function of the various process parameters in Fig.~\ref{fig:5}. Increasing the capillary number  $\widehat{\mathrm{Ca}}$ leads to smaller $\widehat{V}_0^{\mathrm{Cr}}$ and abrupter transition value (a), a larger modified Debye length $\widehat{\lambda}_\mathrm{D}$ increases the critical flux (b) and larger values of the scaled colloidal stress $\widehat{\mathrm{A}}_\mathrm{H}$ decrease the critical flux.
\section{Conclusions and outlook}
Understanding the interaction of droplets with the surface of separation membranes is crucial for developing better materials and improved process conditions aimed at reducing or reversing fouling during oil/water emulsion separation. 
With the use of a hydrodynamic model, coupled with the equation governing the droplet shape, and incorporating colloidal attractive and repulsive stresses, we have shown the existence of different equilibria regions: stable, bistable and unstable. 
These have implications towards regimes under which deposition always occurs, vs. conditions which may reduce the rate of deposition, or possibly increase its reversibility. 
The stability threshold is given by a `critical' scaled permeation $\widehat{V}_0^{\mathrm{Cr}}$ for which, at larger values of the scaled permeation $\widehat{V}_0$, a stable equilibrium ceases to exist. 
An equilibrium phase diagram was constructed in terms of different process parameters, reflecting the relative importance of hydrodynamic and colloidal stresses, both in terms of their magnitude but also in terms of their range. 
Within the phase diagram, stable and unstable droplet shapes are identified. 
Stable droplet shapes are found for to feature upward deflection, due to the prevalence of long--range repulsion -- increasing the electrostatic decay length $\widehat{\lambda}_\mathrm{D}$ results in an increased critical flux $\widehat{V}_0^{\mathrm{Cr}}$. 
Increasing the modified capillary number $\widehat{\mathrm{Ca}}$, representing a more easily deformable droplet, is a primary reason for an increased stable region owing to an upward deflection, but will eventually lead to a lower critical flux and an abrupt transition leading to deposition. 
The scaled colloidal stress $\widehat{\mathrm{A}}_\mathrm{H}$ in turn decreases the critical permeation, making the system less stable. 
Future possible directions stemming from this work are the extension of the model to allow droplet spreading and identifying final shape -– contact area, as well as calculating energy barriers from dynamical simulations.
\begin{acknowledgement}
The research was funded by the Israel Science Foundation (ISF) grant number 2018/17. M. Galvagno was supported in part by a fellowship from the Lady Davis Foundation at the Technion.
\end{acknowledgement}

%\begin{suppinfo}
%Animation of deflection $\hat{d}(\eta)$ as a function of the modified permeation $\widehat{V}_0$. 
%\end{suppinfo}

%\clearpage
\bibliography{GalvagnoRamon20}

\providecommand{\latin}[1]{#1}
\makeatletter
\providecommand{\doi}
  {\begingroup\let\do\@makeother\dospecials
  \catcode`\{=1 \catcode`\}=2 \doi@aux}
\providecommand{\doi@aux}[1]{\endgroup\texttt{#1}}
\makeatother
\providecommand*\mcitethebibliography{\thebibliography}
\csname @ifundefined\endcsname{endmcitethebibliography}
  {\let\endmcitethebibliography\endthebibliography}{}
\begin{mcitethebibliography}{26}
\providecommand*\natexlab[1]{#1}
\providecommand*\mciteSetBstSublistMode[1]{}
\providecommand*\mciteSetBstMaxWidthForm[2]{}
\providecommand*\mciteBstWouldAddEndPuncttrue
  {\def\EndOfBibitem{\unskip.}}
\providecommand*\mciteBstWouldAddEndPunctfalse
  {\let\EndOfBibitem\relax}
\providecommand*\mciteSetBstMidEndSepPunct[3]{}
\providecommand*\mciteSetBstSublistLabelBeginEnd[3]{}
\providecommand*\EndOfBibitem{}
\mciteSetBstSublistMode{f}
\mciteSetBstMaxWidthForm{subitem}{(\alph{mcitesubitemcount})}
\mciteSetBstSublistLabelBeginEnd
  {\mcitemaxwidthsubitemform\space}
  {\relax}
  {\relax}

\bibitem[Mondal and Wickramasinghe(2008)Mondal, and Wickramasinghe]{Mondal2008}
Mondal,~S.; Wickramasinghe,~S.~R. {Produced water treatment by nanofiltration
  and reverse osmosis membranes}. \emph{Journal of Membrane Science}
  \textbf{2008}, \emph{322}, 162--170\relax
\mciteBstWouldAddEndPuncttrue
\mciteSetBstMidEndSepPunct{\mcitedefaultmidpunct}
{\mcitedefaultendpunct}{\mcitedefaultseppunct}\relax
\EndOfBibitem
\bibitem[Vengosh \latin{et~al.}(2014)Vengosh, Jackson, Warner, Darrah, and
  Kondash]{Vengosh2014}
Vengosh,~A.; Jackson,~R.~B.; Warner,~N.; Darrah,~T.~H.; Kondash,~A. {A Critical
  Review of the Risks to Water Resources from Unconventional Shale Gas
  Development and Hydraulic Fracturing in the United States}.
  \emph{Environmental Science {\&} Technology} \textbf{2014}, \emph{48},
  8334--8348\relax
\mciteBstWouldAddEndPuncttrue
\mciteSetBstMidEndSepPunct{\mcitedefaultmidpunct}
{\mcitedefaultendpunct}{\mcitedefaultseppunct}\relax
\EndOfBibitem
\bibitem[Fakhru’l-Razi \latin{et~al.}(2009)Fakhru’l-Razi, Pendashteh,
  Abdullah, Biak, Madaeni, and Abidin]{Ahmadunetal09-jhm}
Fakhru’l-Razi,~A.; Pendashteh,~A.; Abdullah,~L.~C.; Biak,~D. R.~A.;
  Madaeni,~S.~S.; Abidin,~Z.~Z. Review of technologies for oil and gas produced
  water treatment. \emph{Journal of Hazardous Materials} \textbf{2009},
  \emph{170}, 530 -- 551\relax
\mciteBstWouldAddEndPuncttrue
\mciteSetBstMidEndSepPunct{\mcitedefaultmidpunct}
{\mcitedefaultendpunct}{\mcitedefaultseppunct}\relax
\EndOfBibitem
\bibitem[Shaffer \latin{et~al.}(2013)Shaffer, Arias~Chavez, Ben-Sasson,
  Romero-Vargas~Castrill\'on, Yip, and Elimelech]{Shafferetal13-envscitech}
Shaffer,~D.~L.; Arias~Chavez,~L.~H.; Ben-Sasson,~M.;
  Romero-Vargas~Castrill\'on,~S.; Yip,~N.~Y.; Elimelech,~M. Desalination and
  Reuse of High-Salinity Shale Gas Produced Water: Drivers, Technologies, and
  Future Directions. \emph{Environmental Science \& Technology} \textbf{2013},
  \emph{47}, 9569--9583, PMID: 23885720\relax
\mciteBstWouldAddEndPuncttrue
\mciteSetBstMidEndSepPunct{\mcitedefaultmidpunct}
{\mcitedefaultendpunct}{\mcitedefaultseppunct}\relax
\EndOfBibitem
\bibitem[Tanudjaja \latin{et~al.}(2019)Tanudjaja, Hejase, Tarabara, Fane, and
  Chew]{Tanudjaja2019}
Tanudjaja,~H.~J.; Hejase,~C.~A.; Tarabara,~V.~V.; Fane,~A.~G.; Chew,~J.~W.
  {Membrane-based separation for oily wastewater: A practical perspective}.
  \emph{Water Research} \textbf{2019}, \emph{156}, 347--365\relax
\mciteBstWouldAddEndPuncttrue
\mciteSetBstMidEndSepPunct{\mcitedefaultmidpunct}
{\mcitedefaultendpunct}{\mcitedefaultseppunct}\relax
\EndOfBibitem
\bibitem[Bacchin \latin{et~al.}(2006)Bacchin, Aimar, and Field]{Bacchin2006}
Bacchin,~P.; Aimar,~P.; Field,~R. {Critical and sustainable fluxes: Theory,
  experiments and applications}. \emph{Journal of Membrane Science}
  \textbf{2006}, \emph{281}, 42--69\relax
\mciteBstWouldAddEndPuncttrue
\mciteSetBstMidEndSepPunct{\mcitedefaultmidpunct}
{\mcitedefaultendpunct}{\mcitedefaultseppunct}\relax
\EndOfBibitem
\bibitem[Wang \latin{et~al.}(2005)Wang, Guillen, and Hoek]{Wang2005}
Wang,~S.; Guillen,~G.; Hoek,~E. M.~V. {Direct Observation of Microbial Adhesion
  to Membranes}. \emph{Environmental Science {\&} Technology} \textbf{2005},
  \emph{39}, 6461--6469\relax
\mciteBstWouldAddEndPuncttrue
\mciteSetBstMidEndSepPunct{\mcitedefaultmidpunct}
{\mcitedefaultendpunct}{\mcitedefaultseppunct}\relax
\EndOfBibitem
\bibitem[Zhu \latin{et~al.}(2014)Zhu, Wang, Jiang, and
  Jin]{Zhuetal14-NPGasiamaterials}
Zhu,~Y.; Wang,~D.; Jiang,~L.; Jin,~J. Recent progress in developing advanced
  membranes for emulsified oil/water separation. \emph{NPG Asia Materials}
  \textbf{2014}, \emph{6}, e101--e101\relax
\mciteBstWouldAddEndPuncttrue
\mciteSetBstMidEndSepPunct{\mcitedefaultmidpunct}
{\mcitedefaultendpunct}{\mcitedefaultseppunct}\relax
\EndOfBibitem
\bibitem[Tummons \latin{et~al.}(2016)Tummons, Tarabara, Chew, and
  Fane]{tummonsetal16-jmemsci}
Tummons,~E.~N.; Tarabara,~V.~V.; Chew,~J.; Fane,~A.~G. Behavior of oil droplets
  at the membrane surface during crossflow microfiltration of oil?water
  emulsions. \emph{Journal of Membrane Science} \textbf{2016}, \emph{500}, 211
  -- 224\relax
\mciteBstWouldAddEndPuncttrue
\mciteSetBstMidEndSepPunct{\mcitedefaultmidpunct}
{\mcitedefaultendpunct}{\mcitedefaultseppunct}\relax
\EndOfBibitem
\bibitem[Fux and Ramon(2017)Fux, and Ramon]{fuxetal17-envscitech}
Fux,~G.; Ramon,~G.~Z. Microscale Dynamics of Oil Droplets at a Membrane
  Surface: Deformation, Reversibility, and Implications for Fouling.
  \emph{Environmental Science \& Technology} \textbf{2017}, \emph{51},
  13842--13849, PMID: 29110471\relax
\mciteBstWouldAddEndPuncttrue
\mciteSetBstMidEndSepPunct{\mcitedefaultmidpunct}
{\mcitedefaultendpunct}{\mcitedefaultseppunct}\relax
\EndOfBibitem
\bibitem[Tummons \latin{et~al.}(2017)Tummons, Chew, Fane, and
  Tarabara]{Tummons2017}
Tummons,~E.~N.; Chew,~J.~W.; Fane,~A.~G.; Tarabara,~V.~V. {Ultrafiltration of
  saline oil-in-water emulsions stabilized by an anionic surfactant: Effect of
  surfactant concentration and divalent counterions}. \emph{Journal of Membrane
  Science} \textbf{2017}, \relax
\mciteBstWouldAddEndPunctfalse
\mciteSetBstMidEndSepPunct{\mcitedefaultmidpunct}
{}{\mcitedefaultseppunct}\relax
\EndOfBibitem
\bibitem[Tanudjaja and Chew(2018)Tanudjaja, and Chew]{Tanudjaja2018}
Tanudjaja,~H.~J.; Chew,~J.~W. {Assessment of oil fouling by oil-membrane
  interaction energy analysis}. \emph{Journal of Membrane Science}
  \textbf{2018}, \emph{560}, 21--29\relax
\mciteBstWouldAddEndPuncttrue
\mciteSetBstMidEndSepPunct{\mcitedefaultmidpunct}
{\mcitedefaultendpunct}{\mcitedefaultseppunct}\relax
\EndOfBibitem
\bibitem[Tanudjaja and Chew(2019)Tanudjaja, and Chew]{Tanudjaja2019a}
Tanudjaja,~H.~J.; Chew,~J.~W. {Critical flux and fouling mechanism in cross
  flow microfiltration of oil emulsion: Effect of viscosity and bidispersity}.
  \emph{Separation and Purification Technology} \textbf{2019}, \relax
\mciteBstWouldAddEndPunctfalse
\mciteSetBstMidEndSepPunct{\mcitedefaultmidpunct}
{}{\mcitedefaultseppunct}\relax
\EndOfBibitem
\bibitem[Ramon \latin{et~al.}(2013)Ramon, Huppert, Lister, and
  Stone]{ramonetal13-pof}
Ramon,~G.~Z.; Huppert,~H.~E.; Lister,~J.~R.; Stone,~H.~A. On the hydrodynamic
  interaction between a particle and a permeable surface. \emph{Physics of
  Fluids} \textbf{2013}, \emph{25}, 073103\relax
\mciteBstWouldAddEndPuncttrue
\mciteSetBstMidEndSepPunct{\mcitedefaultmidpunct}
{\mcitedefaultendpunct}{\mcitedefaultseppunct}\relax
\EndOfBibitem
\bibitem[Ramon and Hoek(2012)Ramon, and Hoek]{Ramonetal12-jmemsci}
Ramon,~G.~Z.; Hoek,~E.~M. On the enhanced drag force induced by permeation
  through a filtration membrane. \emph{Journal of Membrane Science}
  \textbf{2012}, \emph{392-393}, 1 -- 8\relax
\mciteBstWouldAddEndPuncttrue
\mciteSetBstMidEndSepPunct{\mcitedefaultmidpunct}
{\mcitedefaultendpunct}{\mcitedefaultseppunct}\relax
\EndOfBibitem
\bibitem[Oron \latin{et~al.}(1997)Oron, Davis, and Bankoff]{oronetal97-rmphys}
Oron,~A.; Davis,~S.~H.; Bankoff,~S.~G. Long-scale evolution of thin liquid
  films. \emph{Reviews of modern physics} \textbf{1997}, \emph{69}, 931\relax
\mciteBstWouldAddEndPuncttrue
\mciteSetBstMidEndSepPunct{\mcitedefaultmidpunct}
{\mcitedefaultendpunct}{\mcitedefaultseppunct}\relax
\EndOfBibitem
\bibitem[Manor \latin{et~al.}(2008)Manor, Vakarelski, Tang, O'Shea, Stevens,
  Grieser, Dagastine, and Chan]{Manoretal08-prl}
Manor,~O.; Vakarelski,~I.~U.; Tang,~X.; O'Shea,~S.~J.; Stevens,~G.~W.;
  Grieser,~F.; Dagastine,~R.~R.; Chan,~D. Y.~C. Hydrodynamic Boundary
  Conditions and Dynamic Forces between Bubbles and Surfaces. \emph{Phys. Rev.
  Lett.} \textbf{2008}, \emph{101}, 024501\relax
\mciteBstWouldAddEndPuncttrue
\mciteSetBstMidEndSepPunct{\mcitedefaultmidpunct}
{\mcitedefaultendpunct}{\mcitedefaultseppunct}\relax
\EndOfBibitem
\bibitem[Manor \latin{et~al.}(2008)Manor, Vakarelski, Stevens, Grieser,
  Dagastine, and Chan]{Manoretal08-lang}
Manor,~O.; Vakarelski,~I.~U.; Stevens,~G.~W.; Grieser,~F.; Dagastine,~R.~R.;
  Chan,~D. Y.~C. Dynamic Forces between Bubbles and Surfaces and Hydrodynamic
  Boundary Conditions. \emph{Langmuir} \textbf{2008}, \emph{24},
  11533--11543\relax
\mciteBstWouldAddEndPuncttrue
\mciteSetBstMidEndSepPunct{\mcitedefaultmidpunct}
{\mcitedefaultendpunct}{\mcitedefaultseppunct}\relax
\EndOfBibitem
\bibitem[Israelachvili(2011)]{israelachvili2011intermolecular}
Israelachvili,~J. \emph{Intermolecular and Surface Forces}; Intermolecular and
  Surface Forces; Elsevier Science, 2011\relax
\mciteBstWouldAddEndPuncttrue
\mciteSetBstMidEndSepPunct{\mcitedefaultmidpunct}
{\mcitedefaultendpunct}{\mcitedefaultseppunct}\relax
\EndOfBibitem
\bibitem[Tabor \latin{et~al.}(2011)Tabor, Wu, Lockie, Manica, Chan, Grieser,
  and Dagastine]{Taboretal11-softmatter}
Tabor,~R.~F.; Wu,~C.; Lockie,~H.; Manica,~R.; Chan,~D. Y.~C.; Grieser,~F.;
  Dagastine,~R.~R. Homo- and hetero-interactions between air bubbles and oil
  droplets measured by atomic force microscopy. \emph{Soft Matter}
  \textbf{2011}, \emph{7}, 8977--8983\relax
\mciteBstWouldAddEndPuncttrue
\mciteSetBstMidEndSepPunct{\mcitedefaultmidpunct}
{\mcitedefaultendpunct}{\mcitedefaultseppunct}\relax
\EndOfBibitem
\bibitem[Chan \latin{et~al.}(2011)Chan, Klaseboer, and
  Manica]{chanetal11-advCollIntSci}
Chan,~D.~Y.; Klaseboer,~E.; Manica,~R. Theory of non-equilibrium force
  measurements involving deformable drops and bubbles. \emph{Advances in
  Colloid and Interface Science} \textbf{2011}, \emph{165}, 70 -- 90, A
  collection of papers presented at XIVth International Conference on Surface
  Forces, Moscow - St. Petersburg, Russia, June 21-27 2010\relax
\mciteBstWouldAddEndPuncttrue
\mciteSetBstMidEndSepPunct{\mcitedefaultmidpunct}
{\mcitedefaultendpunct}{\mcitedefaultseppunct}\relax
\EndOfBibitem
\bibitem[Yiantsios and Davis(1990)Yiantsios, and Davis]{YiantsiosDavis90-jfm}
Yiantsios,~S.~G.; Davis,~R.~H. On the buoyancy-driven motion of a drop towards
  a rigid surface or a deformable interface. \emph{Journal of Fluid Mechanics}
  \textbf{1990}, \emph{217}, 547?573\relax
\mciteBstWouldAddEndPuncttrue
\mciteSetBstMidEndSepPunct{\mcitedefaultmidpunct}
{\mcitedefaultendpunct}{\mcitedefaultseppunct}\relax
\EndOfBibitem
\bibitem[Tabor \latin{et~al.}(2012)Tabor, Grieser, Dagastine, and
  Chan]{taboretal12-jcolIntSci}
Tabor,~R.~F.; Grieser,~F.; Dagastine,~R.~R.; Chan,~D.~Y. Measurement and
  analysis of forces in bubble and droplet systems using AFM. \emph{Journal of
  Colloid and Interface Science} \textbf{2012}, \emph{371}, 1 -- 14\relax
\mciteBstWouldAddEndPuncttrue
\mciteSetBstMidEndSepPunct{\mcitedefaultmidpunct}
{\mcitedefaultendpunct}{\mcitedefaultseppunct}\relax
\EndOfBibitem
\bibitem[Doedel \latin{et~al.}(1991)Doedel, Keller, and
  Kernevez]{Doedeletal91-intjbifchaos}
Doedel,~E.; Keller,~H.~B.; Kernevez,~J.~P. Numerical analysis and control of
  bifurcation problems {(I) B}ifurcation in finite dimensions. \emph{Int. J.
  Bifurcation Chaos} \textbf{1991}, \emph{1}, 493--520\relax
\mciteBstWouldAddEndPuncttrue
\mciteSetBstMidEndSepPunct{\mcitedefaultmidpunct}
{\mcitedefaultendpunct}{\mcitedefaultseppunct}\relax
\EndOfBibitem
\bibitem[Dijkstra \latin{et~al.}(2014)Dijkstra, Wubs, Cliffe, Doedel,
  Dragomirescu, Eckhart, Gelfgat, Hazel, Lucarini, Salinger, Phipps,
  Sanchez-Umbria, Schuttelaars, Tuckerman, and Thiele]{Dijkstraetal14-ccp}
Dijkstra,~H.~A.; Wubs,~F.~W.; Cliffe,~A.~K.; Doedel,~E.; Dragomirescu,~I.~F.;
  Eckhart,~B.; Gelfgat,~A.~Y.; Hazel,~A.; Lucarini,~V.; Salinger,~A.~G.;
  Phipps,~E.~T.; Sanchez-Umbria,~J.; Schuttelaars,~H.; Tuckerman,~L.~S.;
  Thiele,~U. Numerical Bifurcation Methods and their Application to Fluid
  Dynamics: {A}nalysis beyond Simulation. \emph{Commun. Comput. Phys.}
  \textbf{2014}, \emph{15}, 1--45\relax
\mciteBstWouldAddEndPuncttrue
\mciteSetBstMidEndSepPunct{\mcitedefaultmidpunct}
{\mcitedefaultendpunct}{\mcitedefaultseppunct}\relax
\EndOfBibitem
\end{mcitethebibliography}

%% graphical table of contents.
%%%%%%%%%%%%%%%%%%%%%%%%%%%%%%%%%%%%%%%%%%%%%%%%%%%%%%%%%%%%%%%%%%%%%
%
%\begin{tocentry}
%%
%%Some journals require a graphical entry for the Table of Contents.
%%This should be laid out ``print ready'' so that the sizing of the
%%text is correct.
%%
%%Inside the \texttt{tocentry} environment, the font used is Helvetica
%%8\,pt, as required by \emph{Journal of the American Chemical
%%Society}.\begin{figure}[ht!]
%\includegraphics[width=1\textwidth]{graph_toc}%
%%The surrounding frame is 9\,cm by 3.5\,cm, which is the maximum
%%permitted for  \emph{Journal of the American Chemical Society}
%%graphical table of content entries. The box will not resize if the
%%content is too big: instead it will overflow the edge of the box.
%%
%%This box and the associated title will always be printed on a
%%separate page at the end of the document.
%%
%\end{tocentry}

\end{document}